\documentstyle[11pt]{article}
\def\ft#1#2{{\scriptstyle {#1 \over #2}}}

\def\ww3{{$W_3$}}

\def\del{\partial}

\begin{document}
\topmargin 0pt
\oddsidemargin 5mm
\begin{titlepage}
\begin{flushright}
CTP TAMU-5/95\\
Imperial/TP/94-95/21\\
hep-th/9502108\\
\end{flushright}
\vspace{1.5truecm}
\begin{center}
{\bf {\Large Embedding of the Bosonic String into the $W_3$ String}}
\vspace{1.5truecm}

{\large H. L\"u$^*$\footnote{Supported in part by the
U.S. Department of Energy, under grant DE-FG05-91-ER40633},\ \  C.N.
Pope$^{*1}$, K.S. Stelle$^\$ $\footnote{Supported in part by the
Commission of the European Communities under contracts
SC1*--CT91--0674 and SC1*--CT92--0789.} and K.W.
Xu$^{*1}$}
\vspace{1.1truecm}

{$*$\ \ \small Center for Theoretical Physics, Texas A\&M University,
                College Station, TX 77843-4242}

\bigskip

{$\$ $\ \ \small The Blackett Laboratory, Imperial College, Prince Consort
Road,
London SW7 2BZ}\vspace{1.1truecm}

\end{center}

\vspace{1.0truecm}

\begin{abstract}
\vspace{1.0truecm}

       We investigate new realisations of the $W_3$ algebra with arbitrary
central charge, making use of the fact that this algebra can be linearised
by the inclusion of a spin-1 current.  We use the new realisations with
$c=102$ and $c=100$ to build non-critical and critical $W_3$ BRST operators.
Both of these can be converted by local canonical transformations into a
BRST operator for the Virasoro string with $c=28-2$, together with a
Kugo-Ojima topological term. Consequently, these realisations provide
embeddings of the Virasoro string into non-critical and critical $W_3$
strings.

\end{abstract}
\end{titlepage}
\newpage
\pagestyle{plain}
\section{Introduction}

     The properties of a string theory are determined by the realisation of
its underlying world-sheet symmetry algebra.   Owing to the non-linearity of
the $W_3$ algebra, which makes realisations hard to come by, the corresponding
$W_3$ strings have only been extensively studied in the free-scalar
realisations found by Romans \cite{romans}.  In this type of realisation, one
of the scalar fields plays a distinguished r\^ole, and the remainder occur
only in the form of their energy-momentum tensor. It is therefore of interest
to look for new realisations that could be used to build other $W_3$ strings.
Some progress has recently been made in understanding the structure of the
$W_3$ algebra.  It has been shown that if the $W_3$ algebra is
extended to $W_{1+3}$ by including a spin-1 current, then after performing a
non-linear basis change this can be recast into the form of a linear algebra
\cite{ks}. After this basis change, the spin-3 current becomes a null
current, and the linear $W_{1+3}$ algebra is generated by this null current,
an energy-momentum tensor, and the spin-1 current.  The procedure can be
reversed.  If the null current is set to zero, one obtains precisely a
Romans-type realisation of the $W_3$ algebra with the derivative of the
distinguished scalar field replaced by the abstract spin-1 current.
However, it was shown in Ref.\ \cite{ks} that the null current of the $W_{1+3}$
algebra can be also be realised by a non-vanishing parafermionic vertex
operator.   In section 2, we obtain new realisations in which the null
current is realised with ghost-like fields.   The central charge of the
resulting $W_3$ realisations can be arbitrary.  An unusual feature of these
realisations is that the spin-3 current commutes with itself at the
classical level.  The $W_3$ symmetry is effectively absent at the classical
level ({\it i.e.}\ at the level of single contractions in OPEs); it arises
properly as a consequence of quantisation.

     The new realisation takes its simplest form when $c=102$.  In section 3,
we shall use this realisation to build a $W_3$ string.  Since the critical
central charge for the $W_3$ algebra is $c=100$, we shall need to make use of
the non-critical BRST operator for $W_3$ that was constructed in
\cite{blnw}.  In this construction, a nilpotent BRST operator is obtained by
taking two abstract realisations of the $W_3$ algebra, with the requirement
that their total central charge adds up to $c=100$. One of the two
realisations can be viewed as the ``Liouville'' sector and the other as the
``matter'' sector.  We shall use the new realisation of the $W_3$ algebra for
the matter sector. Then we need a $c=-2$ realisation for the Liouville sector.
Interestingly enough, this is the unique value of the central charge for which
the $W_3$ algebra can be realised by a single scalar field.  Alternatively, one
can fermionise this scalar and realise the algebra using a pair of fermions
$(b_1, c_1)$, of spins 1 and 0 respectively.  We shall show that the resulting
BRST operator can be re-expressed, by means of a similarity transformation, as
a BRST operator for a Virasoro string with central charge $c=28-2$, together
with a purely topological Kugo-Ojima term whose only non-trivial operator is
the identity.  Thus, the cohomology of this $W_3$ string is entirely equivalent
to that of the usual Virasoro string. This provides an embedding of the
Virasoro string into a non-critical $W_3$ string.

     Another way to use the $c=102$ realisation to build a nilpotent BRST
operator is to gauge also the spin-1 current, since the ghosts for the
additional current make a $c=-2$ contribution to the central charge.  Once the
spin-1 current is included, the constraints become reducible.  The resulting
nilpotent BRST operator for $W_{1+3}$ is equivalent to that for a Virasoro
string with central charge $c=28-2$, as can be seen again by performing a
similarity transformation.  This implies that the {\it non-critical} $W_3$ BRST
operator for the new $c=102$ realisation can be viewed as equivalent to a {\it
critical} $W_{1+3}$ BRST operator, with the fields $(b_1, c_1)$ of the
Liouville sector of the non-critical BRST operator being reinterpreted as the
ghosts for the spin-1 current of the critical BRST operator.

     In section 4, we shall give the details of how to construct a critical
$W_3$ BRST operator, using the new $W_3$ realisation at $c=100$.   After
making a local similarity transformation, we shall show that the BRST operator
is equivalent to that for the $c=28-2$ Virasoro string.  This gives the
embedding of the Virasoro string into the critical $W_3$ string mentioned
above.  This also proves that, in the  new realisation, the $c=102-2$
non-critical $W_3$ string is equivalent to a $c=100$ critical $W_3$ string.

     It is also of interest to examine the problem of embedding the Virasoro
string into extended string theories in a more general context. In addition to
the embedding into the non-critical $W_3$ string mentioned above, in section 2
and section 3 we shall also examine the embedding into a $W_{2,s}$ string,
whose world-sheet symmetry algebra is generated by a spin-$s$ current in
addition to the energy-momentum tensor. In particular, we shall construct
explicit embeddings of the Virasoro string for the cases $s=1,2,3$.

\section{The linear $W_{1+3}$ algebra and new realisations for $W_3$}

     We begin by reviewing the linearisation of the $W_3$ algebra by the
inclusion of a spin-1 current \cite{ks}.  The operator-product expansion of
the spin-1  current with the spin-2 current $T$ and spin-3
current $W$ of the $W_3$ algebra is uniquely determined by requiring that
the Jacobi identities be satisfied.  The resulting $W_{1+3}$ algebra can
then be turned into a linear algebra by a non-linear invertible basis
change:
\begin{eqnarray}
T_0 &=& T\ ,\nonumber\\
W_0 &=& W + \ft13 (JJJ)+\ft12\, (JT) + {3+2t^2\over 2t} \Big(J\del J
+\ft14 \, \del T + {(3+2t^2)\over 6t} \del^2 J\Big)\ ,\label{linearise}\\
J_0 &=& J\ ,\nonumber
\end{eqnarray}
where the $T,W$ currents generate the $W_3$ algebra with central charge
\begin{equation}
c= 50 + 16 t^2 + {36\over t^2} \ ,\label{candc1}
\end{equation}
and the OPEs $J(z)\, W(w)$ and $J(z)\, T(w)$ are uniquely determined by the
Jacobi identities \cite{ks}.  (For simplicity, we have rescaled the $W$
current relative to the usual normalisation.) In this new basis, the OPEs of
the $W_{1+3}$ algebra become linear:
\begin{eqnarray}
T_0(z)\, T_0(0) &\sim& {c\over 2 z^4} + {2T_0 \over z^2} + {\del T_0 \over
z}\ ,\qquad T_0(z)\, W_0(0)\sim {3W_0 \over z^2} + {\del W_0 \over z}\ ,
\nonumber \\
T_0(z)\, J_0(0) &\sim& -{3/t + 2t \over z^3} + {J_0\over z^2} +
{\del J_0\over z}\ , \qquad J_0(z)\, J_0(0)\sim -{1\over z^2}\ ,
\label{w123alge}\\
J_0(z)\, W_0(0)&\sim& {t W_0\over z}\ , \quad W_0(z)\, W_0(0)\sim 0\ .
\nonumber
\end{eqnarray}
Note that in the linear basis, the spin-3 current has become the null
current $W_0$.  The invertibility of the non-linear basis change
(\ref{linearise}) implies that we can reverse the procedure, and obtain a
realisation of the $W_3$ algebra with central charge (\ref{candc1}) in terms
of $W_0, T_0$ and $J_0$. If the null current $W_0$ is set to zero, one
obtains a Romans-type realisation of the $W_3$ algebra, with the derivative
of the distinguished scalar replaced by the abstract spin-1 current $J_0$.
If the null current $W_0$ is non-zero, however, one obtains a new
realisation.   The non-vanishing current $W_0$ was first realised in Ref.\
\cite{ks} using parafermionic vertex operators.   In this paper, we shall
instead realise the $W_0$ current with ghost-like fields.  We first consider
such a realisation of the $W_{1+3}$ algebra (\ref{w123alge}), given by
\begin{equation}
T_0 = T_X + 2 \del r\, s + 3 r\del s\ ,\quad W_0 = r\ ,
\quad J_0 = rs\ ,\label{w123real}
\end{equation}
where $(r,s)$ are a pair of bosonic ghost-like fields with the OPE $r(z)\, s(w)
\sim 1/(z-w)$.  It is easy to see from the $J_0(z)\, W_0(w)$ OPE that $t=-1$.
It follows from Eqn.\ (\ref{candc1}) that $T_0$ must have central charge
$c=102$, and hence from Eqn.\ (\ref{w123real}) that the central charge for the
energy-momentum tensor $T_X$ is $c_X =28$.

    Having found a realisation of the linear $W_{1+3}$ algebra, we may  now
invert the basis change (\ref{linearise}) to obtain a new realisation of  the
$W_3$ algebra with central charge $c=102$:
\begin{eqnarray}
T&=&T_X + 2 \del r\, s + 3 r\, \del s\ ,\nonumber\\
W&=& r - \ft13 r^3s^3 - \ft12 rs\, T_X + \ft{5}2 \del r\,rs^2 +
    \ft18 r\del^2 s + \ft18 \del r\, \del s - \ft{7}4 \del^2 r\, s +
    \ft{5}8 \del T_X\ .\label{newrealise}
\end{eqnarray}
An unusual feature of this realisation, which contrasts with the situation  in
the usual free-scalar realisations, is that the classical terms in the $W$
current, {\it  i.e.} $r -\ft13 r^3s^3$, commute with themselves at the
classical level.\footnote{In this paper, we assign the dimension of the fields
$(b, c, b_1,  c_1, \phi, X^\mu)$ to be $(\hbar, \hbar^0, \hbar, \hbar^0,
\hbar^{\ft12}, \hbar^{\ft12})$ respectively.  For the $(r,s)$ and $(\beta,
\gamma)$ fields,  whose spins $(j,1-j)$ change with context, we assign the
dimensions to be $(\hbar^j, \hbar^{1-j})$.}  Thus, at the classical level, the
realisation (\ref{newrealise}) generates just a Virasoro algebra together with
a null spin-3 current.  It is only at the quantum level that the $W_3$ symmetry
properly manifests itself.  Note that, although the realisation
(\ref{w123real}) for $W_{1+3}$ is reducible, since $J_0 = sW_0$, the
corresponding realisation (\ref{newrealise}) of $W_3$ is not reducible. It is
worth remarking that $T$ and $W$ still generate the $W_3$ algebra with $c=102$
even when the first term, $r$, is absent from $W$.  In that case, one has a
Romans-type realisation, with the derivative of the distinguished scalar
replaced by $J_0=rs$.  If we bosonise the $(r,s)$ fields, $r=\del \xi\,
e^{-\phi}$ and
$s=\eta\, e^{\phi}$, then the $(\xi, \eta)$ fields and $T_X$ appear in the $W$
current in (\ref{newrealise}) only via the energy-momentum tensor $T$ in
Eqn.\ (\ref{newrealise}).  Then one has precisely a Romans realisation, with
$\phi$ as the distinguished scalar together with an effective energy-momentum
tensor $T_X + T_{\xi\eta}=T_X - \eta\del\xi$.  The field $r$ can have
arbitrary spin in this case with the $r$ term absent from $W$, {\it i.e.}\
when $W_0=0$. The inversion of the basis change (\ref{linearise}) then gives
rise to a realisation of the $W_3$ algebra at arbitrary central
charge.\footnote{One can also bosonise the
$(r,s)$  fields differently, $r=e^{i\rho + \phi}$ and $s=-i \del\rho e^{-i\rho
-\phi}$.  In that case, the $W_0$ current is a pure exponential, which is
equivalent to a special case of the parafermionic realisations discussed in
Ref.\ \cite{ks}.}

     The new formulation of the $W_3$ string that we present in this paper is
based on a realisation including the $r$ term in the $W$ current.  The fact
that the realisation (\ref{newrealise}) has central charge $c=102$ is a
consequence of our choice of the realisation (\ref{w123real}) for $W_{1+3}$.
We can construct new realisations, which generalise the one given in Eqn.\
(\ref{w123real}),  to obtain new realisations of $W_3$  with arbitrary
central charge.  To do this, we introduce a pair of  ghost-like
anti-commuting fields $(b_1, c_1)$ with spin $(k,1-k)$. The realisations of
$W_{1+3}$ with arbitrary central charge take the form
\begin{eqnarray}
T_0&=&T_X +2 \del r\, s + 3r\del s -k\, b_1\del c_1 - (k-1)\, \del b_1\,
c_1\ ,\nonumber\\
W_0&=& r\ ,\label{w123realgen}\\
J_0&=&-t\, rs + \sqrt{t^2 -1}\, b_1c_1\ ,\nonumber
\end{eqnarray}
where $k=\ft12 + \ft32t^{-1}\sqrt{t^2-1}$.  Note that the $(b_1,c_1)$ fields
here can be bosonised and hence replaced by a free scalar. The central charge
of the realisation is given by (\ref{candc1}).  When $t^2 =1$, the $b_1$ and
$c_1$ fields do not appear in $J_0$, the central charge is $c=102$, and
$k=\ft12$.  This gives rise to the $W_{1+3}$ realisation (\ref{w123real}), with
$T_X$ replaced by $T_X + T_{b_1c_1}$.  If one chooses $t^2 = \ft98$ instead,
the central charge is $c=100$ and $k=1$.

      Having obtained the realisations (\ref{w123realgen}) of the $W_{1+3}$
algebra, it is straightforward to obtain realisations of $W_3$ with arbitrary
central charge simply by inverting the basis change (\ref{linearise}).  In
particular, when one chooses the central charge parameter $t^2 =\ft98$, one
obtains a realisation of the $W_3$ algebra at $c=100$.  In section 4, we shall
construct the critical $W_3$ BRST operator for this realisation and study its
cohomology. Before doing so, we shall construct in section 3 the non-critical
BRST operator for the realisation (\ref{newrealise}) with $c=102$.

\section{The non-critical $W_3$ BRST operator and its cohomology}

    In this section, we use the $c=102$ realisation (\ref{newrealise})
to construct a $W_3$ string. To do this, we must combine it with a $c=-2$
realisation to form a non-critical BRST operator as constructed in
\cite{blnw}. We shall use a pair of fermions $(b_1, c_1)$  with spins $(1, 0)$
to form the $c=-2$ realisation of $W_3$ \cite{lpsw}:
\begin{equation}
T_{\rm L} = - b_1\del c_1\ , \qquad
W_{\rm L} = \del b_1\, \del c_1 - b_1\,\del^2 c_1\ .\label{wm2real}
\end{equation}
Note that this realisation is reducible, and that it does not close
classically but it does close at the quantum level.  This reducibility is
harmless, however, since this realisation enters only in the Liouville sector
of the theory. By a local canonical field redefinition, the BRST operator can
be transformed into a graded form \cite{bbprs}
\begin{eqnarray}
Q &=& Q_0 + Q_1\ ,\nonumber\\
Q_0&=& \oint c\Big( T -b_1\del c_1 - 2 \del\beta\, \gamma -
3 \beta\del\gamma - b\del c\Big) \ , \label{w3brst1}\\
Q_1&=& \oint \gamma\Big(W +\ft32b_1c_1\, T -\ft92 b_1 c_1\, \beta\del\gamma
-38\del T\nonumber\\
&&+\ft{21}4 b_1\del^2 c_1 + 6 \del b_1\,\del c_1 + \ft{15}4 \del^2 b_1\,
c_1 + \ft34\del \beta\, \del\gamma\Big) \ ,\nonumber
\end{eqnarray}
where $(c,b)$ and $(\gamma, \beta)$ are ghosts and anti-ghosts for the  spin-2
and spin-3 currents respectively, and $T$ and $W$ are given in  Eqn.\
(\ref{newrealise}).  The effective energy-momentum tensor $T_X$ that  appears
in the operator $Q_1$ can be removed by a similarity  transformation
$Q\longrightarrow e^R\,Q\, e^{-R}$, with $R$ given by
\begin{equation}
R = \oint \ft32 b(\gamma b_1 c_1 + \gamma r s + \ft32 \del\gamma)\ .
\end{equation}
The BRST operator $Q$ in (\ref{w3brst1}) can then be re-expressed as
\begin{eqnarray}
Q &=& Q_0 + Q_1\ ,\nonumber\\
Q_0&=& \oint c\Big( T_X + 2\del r\, s + 3 r\del s - b_1\del c_1
- 2 \del\beta\, \gamma -
3 \beta\del\gamma - b\del c\Big) \ , \label{w3brst2}\\
Q_1&=& \oint \gamma\Big(r -\ft13  r^3s^3 +\ft{7}2 \del r\, rs^2 +\ft32
r^2\del s\, s  - \ft12 rsb_1\del c_1  - \ft32 rs\beta\del\gamma
\nonumber\\
&& - \ft{9}4\del^2 r\, s - \del r\, \del s + \ft14 r\del^2 s +
\ft12 b_1\del^2 c_1 +\ft34 \del b_1\, \del c_1 +\ft{5}4 \del\beta\,
\del\gamma\Big)\ .\nonumber
\end{eqnarray}
It is interesting to note that the Liouville sector $(b_1, c_1)$ enters the
operator $Q_1$ only as a quantum correction to the classical terms $\gamma
(r-\ft13 r^3s^3)$.  The BRST operator $Q$ remains nilpotent even if the term
$\gamma r$ is absent.  If one were to bosonise the $(r, s)$ fields in that
case, the BRST operator would become the same as the one given in Ref.\
\cite{bbr}, except that in this paper we use Eqn.\ (\ref{wm2real}) rather than
a two-scalar realisation for the $c=-2$ Liouville sector, and use $T_X +
T_{\xi\eta}$ for the $c=26$ effective energy-momentum tensor.  However, as we
shall see later, the inclusion of the $\gamma r$ term is crucial for obtaining
a genuine embedding of the Virasoro string into the non-critical $W_3$ string.

      The BRST operator (\ref{w3brst2}) for the $W_3\equiv W_{2,3}$ case can
easily be generalised to the cases of $W_{2,1}$ and $W_{2,2}$ strings.  For
the $W_{2,1}$ string, the BRST operator is given by
\begin{eqnarray}
Q_0&=& \oint c\Big( T_X + r\del s -b_1\del c_1 -\beta\del\gamma -
b\del c\Big)\ ,\nonumber\\
Q_1&=& \oint \gamma (r - r s - b_1 c_1)\ .\label{w21brst}
\end{eqnarray}
For the $W_{2,2}$ case, it is given by
\begin{eqnarray}
Q_0&=& \oint c\Big( T_X + \del r\, s + 2r\del s -b_1\del c_1 -\del
\beta\, \gamma -2\beta\del\gamma - b\del c\Big)\ ,\nonumber\\
Q_1&=& \oint \gamma (r -\ft12 r^2s^2 +  r\del s + 2 \del r\, s -
b_1\del c_1 -\beta\del\gamma)\ .\label{w22brst}
\end{eqnarray}
There should be no confusion from the fact that the $(r, s)$ and $(\beta,
\gamma)$ fields have different conformal spins in different BRST operators.
It is easy to  verify that the $(r, s)$ terms in $\{Q_1, \beta\}$, for $Q_1$
given in (\ref{w22brst}), realise the Virasoro algebra with central charge
$c=28$.

    Having constructed the non-critical BRST operator for the $c=102$
realisation (\ref{newrealise}) of $W_3$,  we now would like to study its
cohomology. In fact, we shall show that the BRST operator given in
(\ref{w3brst2}) provides an embedding of the Virasoro string into the
non-critical $W_3$ string.  As we shall see later, the BRST operators given
in (\ref{w21brst}), (\ref{w22brst}) and (\ref{w3brst2}) can all be
re-expressed, by local similarity transformations, in the form
\begin{eqnarray}
Q_0&=& \oint c\Big( T_X + (j-1)\del r\, s + j\, r\del s - b_1\del c_1 -
(j-1)\del \beta\, \gamma - j\, \beta\del \gamma - b\del c\Big)
\ ,\nonumber\\
Q_1&=& \oint \gamma r \ ,\label{brst123}
\end{eqnarray}
for $j=1,2,3$ respectively.  This BRST operator can be simplified further by
the similarity transformation $Q \longrightarrow e^R Q e^{-R}$, with $R$
given by
\begin{equation}
R = \oint s(c\del \beta + j\, \del c \, \beta)\ ,
\end{equation}
leading to
\begin{equation}
Q_0 = \oint c\Big( T_X -b_1\del c_1 -b \del c\Big) \ , \qquad
Q_1 = \oint \gamma r\ .\label{brsttop}
\end{equation}
Since the $(\beta, \gamma)$ and $(r, s)$ fields form a Kugo-Ojima quartet
\cite{kj}, the  BRST operator (\ref{brsttop}) simply describes a $c=26$
dimensional Virasoro string with an energy-momentum tensor given by $T_X
-b_1\del c_1$.

     In the above similarity transformations, the most difficult  part is the
transformation of the BRST operators (\ref{w21brst}),  (\ref{w22brst}) and
(\ref{w3brst2}) into the form given in (\ref{brst123}). We shall discuss this
procedure case by case.  First, let us consider the $W_{2,1}$ BRST operator
given in (\ref{w21brst}).  The similarity transformation involves an infinite
number of terms, given by
\begin{equation}
R =\sum_{n\ge 1} R_n = -\oint \sum_{n\ge 1} {2^{1-n}\over n(n+1)}
\Big(r s^{n+1} + (n+1) b_1 c_1 s^n\Big)\ .\label{rforw21}
\end{equation}
It is easy to see that this similarity transformation converts the BRST
operator (\ref{w21brst}) into the form (\ref{brst123}) with $j=1$.  Since
$R$ is primary under the total energy-momentum tensor, and since it does not
contain the $(b,c)$ ghosts, this transformation leaves the operator $Q_0$
invariant.  That $Q_1$ is converted into the form $\gamma r$ is a consequence
of the fact that $e^{-R} r e^R =r - rs - b_1c_1$.  Note that all the terms in
Eqn.\ (\ref{rforw21}) are classical.  This is a special feature for the
$W_{2,1}$ case, which does not occur for higher-spin cases.  The absence of
quantum corrections here occurs simply because they are all total derivatives
when
$j=1$.

     Next we consider the $W_{2,2}$ case, whose BRST operator is given in
(\ref{w22brst}).  In this case, the full similarity transformation is very
complicated.  It is instructive to study the transformation first at the
classical  level.  The classical terms of the $Q_1$ operator are $\gamma(r -
\ft12  r^2s^2)$.  At the classical level, there is an invertible local
canonical transformation of the $(r,s)$ fields that converts the classical
terms into $\gamma r$, namely
\begin{eqnarray}
r&\longleftarrow& r -\ft12 r^2\, s^2\ ,\nonumber\\
s&\longleftarrow& \sum_{n\ge 0} g_n\, r^{n}\, s^{2n+1}\ ,\label{w22red}
\end{eqnarray}
where $g_n = n(2n+1)^{-1}\, g_{n-1}$ with $g_0=1$.  It is easy to verify  that
this canonical transformation leaves the associated energy-momentum tensor
appearing in the $Q_0$ operator (\ref{w22brst}) invariant at the  classical
level.  The classical terms of the generator $R$ associated with the
transformation (\ref{w22red}) are more complicated. They take the form
\begin{equation}
R_{\rm cl} = \sum_{n\ge 1} R_n = \oint \sum_{n\ge 1} h_n\, r^{n+1}\,
s^{2n+1}\ .\label{rforw22cl}
\end{equation}
The first few coefficients are given by $h_1=\ft16$, $h_2=\ft1{60}$, $h_3
=\ft{41}{15120}$, $h_4=\ft1{1890}$, $h_5=\ft{3337}{29937600}$, {\it etc.}  In
order to obtain the generator $R$ at the full quantum level, it is
also necessary to add quantum corrections to the classical terms
(\ref{rforw22cl}).  Although the $R$ operator involves infinitely many terms,
it is possible to construct it in a systematic way since the terms can be
organised according to their grading degree under the $(r,s)$ number operator
$J = rs$.  In fact, the classical terms of $R_n$ in Eqn.\ (\ref{rforw22cl})
have degree $n$.  For each
$R_n$, there is only a finite number of candidates for possible quantum
corrections. Since the similarity transformation leaves $Q_0$ invariant, $R_n$
has to satisfy $[ R_n, T^{\rm tot} ]=0$, with $T^{\rm tot} = \{Q_0, b\}$.
Furthermore, it has to satisfy the following recursive conditions organised by
the $(r,s)$ grading:
\begin{eqnarray}
{[R_1, \gamma r]} &=& Q_1' \ ,\nonumber\\
{[ R_2, \gamma r ]} -\ft12 {[ R_1, [ R_1, \gamma r ]]}&=& 0\ ,
\label{w22recur}\\
{[ R_3, \gamma r ]} -\ft12 {[ R_1, [ R_2, \gamma r ]]} -\ft12
{[R_2, [R_1, \gamma r]]} + \ft16 {[R_1, [R_1, [R_1, \gamma r]]]} &=& 0
\ ,\nonumber\\
\cdots\cdots&& \nonumber
\end{eqnarray}
where $Q'_1$ is $Q_1$ in Eqn.\ (\ref{w22brst}) with the $\gamma r$ term
removed.  The computation can be simplified by noting that the classical
terms in $R_n$ also have degree zero under  the $(b_1, c_1)$ and $(\beta,
\gamma)$ number operators.  By making the  ansatz that this property persists
at the quantum level, the number of  candidate terms for each $n$ is greatly
reduced.  $R_1$ can easily be solved, and is given by
\begin{equation}
R_1 = \oint s(\ft16 r^2\, s^2 +\ft32 r\del s + \beta\del \gamma + \ft12
\del\beta\, \gamma +  b_1\del c_1)\ .
\end{equation}
Note that the general forms of the quantum corrections can be obtained by
making the  replacements $rs \rightarrow \del$, or $r^2s^2 \rightarrow b_1
c_1\del$, {\it etc.}  The numbers of possible quantum corrections in $R_n$ and
the complexity of the computations increase rapidly with $n$.  We have  solved
the recursive conditions (\ref{w22recur}) explicitly for $R_1, R_2$ and
$R_3$. It is worth remarking that all the coefficients in $R_1$ are fixed by
the recursive conditions (\ref{w22recur}). There are two free parameters in
$R_2$ and 10 free parameters in $R_3$.  Although we have not completely
analysed the recursive conditions (\ref{w22recur}) for higher values of $n$,
we expect that the solutions exist and that the number of free parameters
increases with $n$.  The fact that solutions of this form can be found
justifies the simplifying ansatz made above.

      Having obtained the similarity transformations for the $W_{2,1}$ and
$W_{2,2}$ cases, it is straightforward to discuss the $W_{2,3}$ case.  The
similarity transformation generator here is even more complicated again.  As in
the  cases discussed above, it is instructive first to study the
transformation at the classical level.  The classical terms of the  operator
$Q_1$ given in (\ref{w3brst2}) are $\gamma(r -\ft13 r^3s^3)$. They can be
transformed classically into $\gamma r$ by the invertible local field
redefinition
\begin{eqnarray}
r&\longleftarrow& r -\ft13 r^3\, s^3\ ,\nonumber\\
s&\longleftarrow& \sum_{n\ge 0} g_n\, r^{2n}\, s^{3n+1}\ ,
\label{w23red}
\end{eqnarray}
where $g_n = n(3n+1)^{-1} g_{n-1} $ and $g_0 =1$.  This field redefinition
leaves the operator $Q_0$ in (\ref{w3brst2}) invariant.  The terms of the
similarity transformation generator $R$ for the classical transformation
(\ref{w23red}) take the form
\begin{equation}
R_{\rm cl} = \sum_{n\ge 1} R_n =\oint \sum_{n\ge 1}
h_n\, r^{2n+1}\, s^{3n+1}\ ,\label{rforw23cl}
\end{equation}
where $h_1 =\ft1{12}$, $h_2 = \ft1{168}$, $h_3 = \ft{43}{60480}$, $h_4 =
\ft{41}{393120}$, $h_5 = \ft{26917}{1585059840}, etc$.   Analogously to  the
$W_{2,2}$ case, the $R_n$ term in Eqn.\ (\ref{rforw23cl}) has a grading degree
$n$ under the $(r,s)$ number operator $J=rs$, and has degree zero under the
$(b_1,c_1)$ and $(\beta, \gamma)$ number operators.  Assuming this grading
structure persists at the quantum level, we can consider the possible quantum
corrections to $R_n$.  In addition to the requirement that
$[ R_n, T^{\rm tot}] = 0$, with $T^{\rm tot} = \{Q_0,b\}$, $R_n$ has to
satisfy the recursive conditions organised by the $(r, s)$ grading, as given
in (\ref{w22recur}), where $Q'_1$ is now the operator $Q_1$ given in
(\ref{w3brst2}) with the $\gamma r$ term absent.  The complexity of solving
these recursive conditions increases dramatically with $n$.  We have solved
explicitly for $R_1$ and $R_2$.  As in the $W_{2,2}$ case, the number of  free
parameters in $R_n$ increases with $n$.

     While we have not given in closed form the full quantum-level
canonical transformations needed to cast the BRST operator into the form of
Eqn.\ (\ref{brst123}), some aspects of the closed-form expression can
nevertheless be outlined. At the classical level, the transformation
(\ref{w23red}) maps $(r,s)$ to $(\tilde r,\tilde s)$, where
\begin{eqnarray}
\tilde r&=&r(1-\ft13x),\nonumber\\
\tilde s&=&sg(x),\label{rstilde}
\end{eqnarray}
where $x=r^2s^3$. The transformation (\ref{rstilde}) is canonical provided the
Poisson brackets for $r$ and $s$ are preserved, which requires that the
function $g(x)$ satisfy the differential equation
\begin{equation}
x(3-x){d\over dx}g(x) + (1-x)g(x)=1.\label{gdiff}
\end{equation}
This differential equation can be solved:
\begin{equation}
g(x) = {k + \int^x dy\,y^{-2/3}(1-\ft13y)^{-1/3}\over
3x^{1/3}(1-\ft13x)^{2/3}},\label{gsol}
\end{equation}
where $k$ is an integration constant. The integral in (\ref{gsol}) yields a
hypergeometric function. Methods for promoting such classical canonical
transformations to the quantum level in $d=1$ quantum mechanical systems are
now well developed \cite{qcan}, and one may anticipate the applicability of
such methods to the quantum forms of transformations like Eqns.\
(\ref{w22red}) and (\ref{w23red}) in $d=1+1$ conformal field theory as well.

     To end this section, we recall that Eqn.\ (\ref{w123real}) gives a
critical  realisation of the $W_{1+3}$ algebra (\ref{w123alge}).  Thus, one
can  construct a nilpotent BRST operator for the system.  The constraints are
then reducible, however, since $J_0 = s W_0$, and therefore the spin-1 gauge
symmetry in this system is effectively fictitious.  This can also be seen from
the BRST operator:
\begin{eqnarray}
Q&=& \oint \Big( c(T_0 - 3\beta\del \gamma -2 \del\beta\, \gamma -
b_1\del c_1 - b\del c) + \gamma W_0 + c_1 J_0 - c_1\beta\gamma
\Big)\ .\label{w123brst}
\end{eqnarray}
After inserting the realisation (\ref{w123real}), one can use a similarity
transformation with $R= \oint s(c\del \beta + 3 \del c\, \beta - \beta c_1)$ to
convert the BRST operator (\ref{w123brst}) into the form of Eqn.\
(\ref{brsttop}), where the ghost fields $(b_1, c_1)$ originally corresponding
to the spin-1 current can now be reinterpreted as a Liouville sector.  Thus,
in this realisation the {\it non-critical} $W_3$ string is equivalent to a
{\it critical} $W_{1+3}$ string.

\section{The critical $W_3$ BRST operator and its cohomology}

     Now we construct a critical BRST operator using the new realisation of
the $W_3$ algebra at $c=100$.  As discussed in section 2,  the new realisation
can be obtained from the $W_{1+3}$ realisation (\ref{w123realgen}) by
inverting  the basis change (\ref{linearise}).   By making a canonical field
redefinition \cite{lpssw}, we find that the corresponding critical BRST
operator can be written in a graded form, given by
\begin{eqnarray}
Q_0 &=& \oint c\Big ( T_X + 2\del r\, s + 3 r\del s -b_1\del c_1
-2\del\beta\, \gamma - 3 \beta\del\gamma -b\del c\Big)\ ,
\nonumber\\ Q_1 &=& \oint \gamma \Big (r - \ft13 r^3s^3 + \ft13 r^2s^2 b_1c_1
+
\ft{10}3 \del r\, rs^2  + \ft43 r^2\del s\, s - \ft23 rs b_1\del c_1
\nonumber\\ &&-\ft89 rs\del b_1\, c_1 - \ft43 rs\beta\del \gamma -\ft{55}{27}
\del^2r\, s -
\ft{20}{27} \del r\, \del s + \ft8{27} r\del^2 s -
\ft{10}{9} \del r\, s b_1 c_1
\label{w3crbrst}\\ &&-\ft49 r\del s\, b_1 c_1 + \ft19 b_1\del^2 c_1 + \ft49
\del b_1\, \del c_1 +
\ft{10}{27} \del^2 b_1\, c_1 + \ft49\beta\del \gamma\, b_1 c_1 +
\ft{28}{27} \del\beta\, \del\gamma\Big)\ ,\nonumber
\end{eqnarray}
where $T_X$ is an energy-momentum tensor with central charge $c=28$.

     In fact, this BRST operator can also be converted into the form of Eqn.\
(\ref{brst123}) by a local similarity transformation.  It then follows that
its cohomology is equivalent to that of the Virasoro string, with an
energy-momentum tensor $T_X -b_1\del c_1$.  Thus, this BRST operator provides
an embedding of the $c=28-2$ Virasoro string into the critical $W_3$ string.
To see this, as in the case of the non-critical $W_3$ BRST operator discussed
in the previous section, it is instructive to study first the similarity
transformation at the classical level.  The classical terms of the $Q_1$
operator (\ref{w3crbrst}) are $\gamma(r -\ft13 r^3 s^3 + \ft13 r^2s^2b_1
c_1)$.  They can be transformed classically into a single term
$\gamma r$ by the invertible local transformation
\begin{eqnarray}
r&\longleftarrow& r - \ft13 r^3s^3 + \ft13 r^2s^2 b_1 c_1 \ ,\nonumber\\
s&\longleftarrow& \sum_{n\ge 0} g_n r^{2n} s^{3n+1}\ ,\nonumber\\
b_1&\longleftarrow& \sum_{n\ge 0} f_n r^{2n}s^{3n} b_1 \ ,
\label{rsbcred}\\
c_1&\longleftarrow& \sum_{n\ge 0} e_n r^{2n}s^{3n} c_1 \ ,\nonumber
\end{eqnarray}
where $g_n = n(3n+1)^{-1} g_{n-1}$, $f_n=\ft19(3n-4)n^{-1} f_{n-1}$, $e_n =
\ft19 (3n-2)n^{-1} e_{n-1}$ and $g_0=f_0=e_0=1$.  This transformation leaves
the operator $Q_0$ in Eqn.\ (\ref{w3crbrst}) invariant.  The terms of the
operator $R$ that generate this classical-level transformation are given by
\begin{equation}
R_{\rm cl} = \sum_{n\ge 1} R_n =
\oint \sum_{n\ge 1} h_n \Big( r^{2n+1}s^{3n+1} - \ft13(3n+1)
r^{2n} s^{3n} b_1 c_1\Big) \ ,\label{rforw23cr}
\end{equation}
where the $h_n$ have precisely the same values as those given below  Eqn.\
(\ref{rforw23cl}).   Analogously to the case of the non-critical $W_3$  BRST
operator, one can then obtain the similarity transformation at the quantum
level, again organised by the $(r,s)$ grading.  Owing to the complexity of the
computation, we have solved explicitly only for $R_1$ and $R_2$.

      In this and the previous sections, we have used the new realisations to
construct critical and non-critical $W_3$ BRST operators.  Both of these
provide embeddings of the $c=28-2$ Virasoro string.  It is also of interest to
compare the properties of the two BRST operators.   First, we note that $Q_0$
is the same for each case.  Also, both BRST operators are nilpotent even in
the absence of the $\gamma r$ term in $Q_1$. When the $\gamma r$ term is
included they are equivalent, since both of them can be converted into
(\ref{brsttop}) by local canonical transformations. When the $\gamma r$ term
is absent, the two BRST operators become inequivalent. Neither of them is then
related to (\ref{brsttop}) by a local canonical transformation. When the
$\gamma r$ term is omitted, it is convenient to consider the BRST operators
with $(r,s)$ and $(b_1,c_1)$ expressed in bosonised form
\begin{eqnarray}
r&=&\del\xi e^{-\phi}\ ,\qquad s=\eta e^{\phi}\ ,\nonumber\\
b_1 &=& e^{-i\rho}\ ,\qquad c_1 = e^{i \rho}\ .
\end{eqnarray}
In terms of the scalar fields $\phi$ and $\rho$, the non-critical BRST
operator becomes the same as the one constructed in \cite{bbr}, except that in
this paper we have realised the $c=-2$ Liouville sector by the single scalar
$\rho$ rather than by a two-scalar system.  The critical BRST operator, on the
other hand, is identical to that of a $c=100$ Romans realisation, with the
distinguished scalar $\phi_1$ given by $(3\phi + i\rho)/\sqrt8$, and the
effective energy-momentum tensor $T^{\rm eff}$ given by $T_X -\ft12
(\del\phi_2)^2 - \sqrt{7/24}\del^2 \phi_2$, where $\phi_2 = (i\phi -
3\rho)/\sqrt8$.  It is easy to see that $T^{\rm eff}$ then generates the
Virasoro algebra with $c=25\ft12$.

      The inequivalence of the two BRST operators with the $\gamma r$ term
omitted can be seen in a more general context by studying the most general
possible forms of the $Q_1$ operator that can be constructed with two scalar
fields.  We have found that, even at the classical level, there are precisely
two inequivalent solutions satisfying the nilpotency conditions.  Both of
these can be extended to the full quantum level.  One corresponds to the
non-critical $W_3$ BRST operator; the other corresponds to the critical $W_3$
BRST operator.  We have showed above that for both of these BRST operators one
can add a term $\gamma r$ after  replacing one of the scalars by an $(r,s)$
system, and with this inclusion the critical and non-critical constructions
become equivalent, and are also equivalent to an embedding of the Virasoro
string.

\section{Conclusions and discussion}

     In this paper, we have constructed a new realisation of the $W_3$
algebra, making use of the fact that the $W_3$ algebra can be linearised
through the addition of a spin-1 current. The spin-3 current $W_0$ in the
linearised $W_{1+3}$ algebra is null.  The previously known free-scalar
realisation of $W_3$ corresponds to the case where this null current is
identically zero.  The null current, however, can be non-zero and was first
realised using parafermionic vertex operators in Ref.\ \cite{ks}. In this
paper the null current is realised by ghost-like fields. The central charge
of the resulting $W_3$ realisation can be arbitrary.  In particular, we
constructed the cases with central charge $c=102$ and $c=100$, and we built
the corresponding non-critical and critical $W_3$ string BRST operators.  We
showed that both BRST operators can be converted into the form of a BRST
operator for the Virasoro string with $c=28-2$, together with a Kugo-Ojima
term.  Thus, the $c=28-2$ Virasoro string can be embedded into both the
critical and the non-critical $W_3$ strings using the new realisation. Since
the Kugo-Ojima system has the identity as its only non-trivial operator, one
may consider these to be genuine embeddings, not requiring any further
truncations of states. The embedding of the $c=26$ Virasoro string into
strings with enlarged worldsheet symmetries was first discussed in Ref.\
\cite{bv}, where its embedding into the $N=1$ and $N=2$ superstrings was
given.

     The embedding of the Virasoro string into the critical $W_3$ string has
also been previously discussed from a different point of view. The standard
Romans realisation of $W_3$ at $c=100$ can be viewed in a loose sense as an
embedding of the $c=25\ft12$ Virasoro string coupled to the Ising model, since
physical states of the $W_3$ string are the tensor product of
$c=25\ft12$ string states with primary operators of the Ising model.  This
description of the embedding is not totally satisfactory, however, since the
$Q_1$ operator, which in this case describes the Ising model, is not purely
topological.  Later, Ref.\ \cite{bfw} proposed an embedding of the $d=26$
Virasoro string into the critical $W_3$ string with the usual Romans
multi-scalar realisation.  However, it had already been established that the
cohomology of the multi-scalar $W_3$ string corresponds to a $c=25\ft12$
bosonic string  coupled to the Ising model, and this is inequivalent to the
cohomology of the  $d=26$ bosonic string.  No local transformation of the
field variables can alter this result.  Indeed, the similarity transformation
used in Ref.\ \cite{bfw} to relate the $W_3$ string to the
$d=26$ Virasoro string involved non-local terms.  Such a transformation does
not establish an equivalence of the cohomologies of two BRST operators, since
the notion of BRST non-triviality must be defined with respect to a given set
of local operators.  The embeddings described in this paper are genuine, in
the sense that they make use only of {\it local} canonical transformations in
order to establish the equivalence of the BRST cohomologies.

     In closing, it is worth remarking that the operator $Q_1$ for the
$W_{2,2}$ string  given in Eqn.\ (\ref{w22brst}) is itself a BRST operator for
the Virasoro  algebra, with $(\beta,\gamma)$ viewed as the ghosts for the
energy-momentum  tensor.  We showed in section 3 that this BRST operator can be
converted into the single term $\gamma r$, for which the only non-trivial
cohomology  is the identity.  Thus, this can be viewed as the BRST operator for
an embedding of an ``empty'' string into the bosonic string.

\section*{Acknowledgements}

     H.L., C.N.P. and K.S.S. would like to thank SISSA for
hospitality at the beginning of this work. K.S.S. would also
like to thank Texas A\&M University for hospitality at its
conclusion.

\section*{Note Added}

     After this paper was completed, we learned of a paper by F.\
Bastianelli and N.\ Ohta \cite{bo}, which has some overlap with our work.
They introduced an additional pair of ghost-like bosonic fields $(\tilde r,
\tilde s)$ which have spins $(1, 0)$, and constructed a $c=102$ realisation
of the $W_{1+3}$ algebra with $W_0=r$ and $J=rs + \tilde r$.  They then
obtained a $c=102$ realisation of $W_3$ and used it to construct a
non-critical $W_3$ BRST operator.  However, they did not obtain the
cohomology of the BRST operator.

\end{document}